\documentclass[conference]{IEEEtran}
\IEEEoverridecommandlockouts
\newtheorem{Remark}{\it Remark}[section]

\newtheorem{Proposition}{\it Proposition}[section]

\makeatletter
\newcommand{\Rmnum}[1]{\expandafter\@slowromancap\romannumeral #1@}
\makeatother

\usepackage{booktabs}
\usepackage{amsmath, amssymb, amsfonts}
\usepackage{graphicx}
\usepackage{bm}
\usepackage{multicol}
\usepackage{setspace}
\usepackage{subfigure}
\usepackage{epstopdf}
\usepackage{fancyhdr} 
\usepackage{indentfirst}
\usepackage{enumerate}
\usepackage{stfloats}
\usepackage{bm}
\usepackage{multirow}
\usepackage{threeparttable}
\usepackage{CJK}
\usepackage[justification=centering]{caption}
\usepackage{makecell}
\usepackage{caption}
\usepackage{cases}
\usepackage{textcomp}
\usepackage{algpseudocode}
\usepackage{amsmath}
\usepackage{graphics}
\usepackage{epsfig}
\usepackage{cite}
\usepackage{color}
\usepackage{caption}

\usepackage{geometry}
\begin{document}
\captionsetup[figure]{labelformat={default},labelsep=period,name={Fig.}}
\title{
Reference Signal-Based Waveform Design for Integrated Sensing and Communications System
}

\author{\IEEEauthorblockN{Ming Lyu, Hao Chen, Dan Wang, Guangyin Feng, Chen Qiu and Xiaodong Xu}
\thanks{
This work was supported in part by the Major Key Project of PCL Department of Broadband Communications under Grant No. PCL2023AS1-1, and in part by the Fundamental Research Funds for the Central Universities under Grant No. 2023ZYGXZR106.

M. Lyu is with the School of Future Technology, South China University of Technology, Guangzhou, China, 511442, and also with the Department of Broadband Communication, Pengcheng Laboratory, Shenzhen, China, 518055. Email: 202211092228@mail.scut.edu.cn.

H. Chen, D. Wang and C. Qiu are with the Department of Broadband Communications, Pengcheng Laboratory, Shenzhen, China, 518055, Emails: chenh03@pcl.ac.cn, wangd01@pcl.ac.cn, qiuch@pcl.ac.cn. 

G. Feng is with the School of Microelectronics, South China University of Technology, Guangzhou, China, 511442. Email: gyfeng88@scut.edu.cn.

X. Xu is with the State Key Laboratory of Networking and Switching Technology, Beijing University of Posts and Telecommunications, Beijing, China, 100876, and also with the Department of Broadband Communication, Pengcheng Laboratory, Shenzhen, China, 518055. Email: xuxiaodong@bupt.edu.cn.
}
}

\maketitle

\begin{abstract}
Integrated sensing and communications (ISAC) as one of the key technologies is capable of supporting high-speed communication and high-precision sensing for the upcoming 6G. This paper studies a waveform strategy by designing the orthogonal frequency division multiplexing (OFDM)-based reference signal (RS) for sensing and communication in ISAC system. We derive the closed-form expressions of Cramér-Rao Bound (CRB) for the distance and velocity estimations, and obtain the communication rate under the mean square error of channel estimation. Then, a weighted sum CRB minimization problem on the distance and velocity estimations is formulated by considering communication rate requirement and RS intervals constraints, which is a mixed-integer problem due to the discrete RS interval values. To solve this problem, some numerical methods are typically adopted to obtain the optimal solutions, whose computational complexity grow exponentially with the number of symbols and subcarriers of OFDM. Therefore, we propose a relaxation and approximation method to transform the original discrete problem into a continuous convex one and obtain the sub-optimal solutions. Finally, our proposed scheme is compared with the exhaustive search method in numerical simulations, which show slight gap between the obtained sub-optimal and optimal solutions, and this gap further decreases with large weight factor.

\end{abstract}

\begin{IEEEkeywords}
Integrated sensing and communications (ISAC), reference signal (RS), orthogonal frequency division multiplexing (OFDM), Cramér-Rao Bound (CRB).
\end{IEEEkeywords}

\section{Introduction}    
With the rapid development of information technology, the sixth-generation (6G) wireless systems are envisioned to not only enhance the communications but also sense the environment to enable seamless interaction between the physical and digital worlds. This capability is crucial for the intelligent applications such as drones, vehicle-to-everything, and even meta-universes \cite{wang_integrated_2022}. Integrated sensing and communications (ISAC), which combine both sensing and communication functions by sharing the hardware and spectrum, aims to improve the spectrum efficiency \cite{wang_receiver_2024, tian_performance_2024}. However, the waveform designs of communication and sensing have different requirements, i.e., communication waveforms are usually designed to transmit random data while sensing waveforms are designed to exhibit constant envelope properties  \cite{hua_mimo_2024}. Hence, to investigate the integrated waveform for both communication and sensing has the potential to further enhance the performance of ISAC systems.

In general, conventional waveform designs are divided into three categories, i.e., communication-centric designs \cite{liyanaarachchi_optimized_2021}, sensing-centric designs \cite{ma_frac_2021}, and joint waveform designs \cite{liu_integrated_2022}. Communication-centric designs aim to incorporate the sensing function into the  existing communication waveform without significant modification, which usually depends on  orthogonal frequency division multiplexing (OFDM) techniques and offers fast-deployment and cost-efficiency\cite{wei_integrated_2023}. On the other hand, Sensing-centric designs focus on embedding communication data into the primary sensing waveform, which may cause limited communication performance \cite{ma_frac_2021}. To balance both communication and sensing requirements, joint waveform designs are recently proposed to conceive an ISAC waveform from the ground-up, instead of relying on existing waveforms  \cite{chen_joint_2021}. The reference signal (RS)-based waveform design is highly compatible with the physical layer of current communication systems and facilitates a smooth transition toward 6G \cite{li_frame_2024}. The RS is known in the modulation domain and is inserted in the OFDM signal for channel estimation, which exhibits good autocorrelation \cite{hsu_analysis_2022}. Consequently, the RS exhibits significant potential for application in radar sensing.

The existing RS in OFDM systems have fixed structures, includes synchronization signal (SS), demodulation RS (DMRS), channel state information-RS (CSI-RS), positioning RS (PRS), and so on \cite{ji_networking_2023}. Specifically, the authors in \cite{cui_integrated_2022} analyzed the sensing performance of SS and DMRS by self-ambiguity and cross-ambiguity functions. On the other hand, the authors in \cite{wei_5g_2023} demonstrated the feasibility and superiority of PRS by comparing with the SS, DMRS and CSI-RS. However, most of the existing related works mainly focus on communication-assisted sensing, whose RS structure is usually fixed, resulting in limited capability and cannot meet the various ISAC requirements. Therefore, it is necessary to design more flexible RS-based waveform strategies in ISAC systems.

This paper fills the gaps in the research of OFDM-based ISAC systems by exploring the design of RS-based waveform strategies to enable simultaneous sensing and communications. To enhance sensing accuracy while maintaining communication requirement, the intervals of RS are designed in the 2D time-frequency domains. Then, the closed-form expression of Cramér-Rao Bound (CRB) for the distance and velocity are derived, and the relationship between communication rate and different RS intervals is calculated by using the mean square error (MSE) of channel estimation. Finally, we formulate a weighted sum CRB minimization problem under the constraints of the communication rate and RS intervals, and propose a relaxation and approximation approach to transform the original non-convex problem into a convex version.

\section{ISAC System Model} 
\begin{figure}[htbp]
    \centering
    \includegraphics[width=0.47\linewidth]{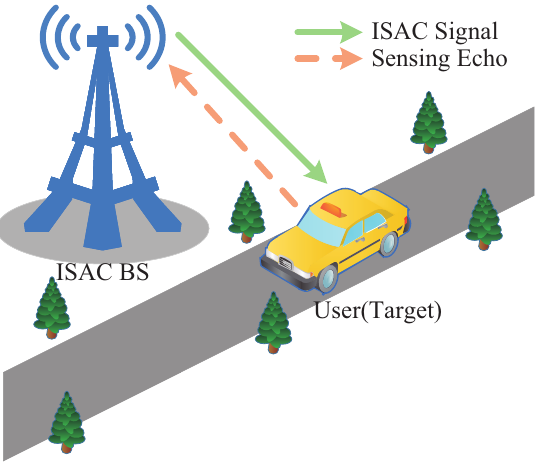}
    \caption{The ISAC system model.}
    \label{fig1}
\end{figure}

We consider an ISAC system with one base station (BS) and a user, as shown in Fig. \ref{fig1}. The user not only acts as a downlink communication device but also acts as a sensing target. The BS acts as an ISAC transceriver to transmit the ISAC signal to user for communications, and then to receive the sensing echo signal reflected by user for sensing. For simplicity, ideal self-interference cancellation is considered in this paper. Here, single antenna is equipped at the BS and user, and multi-antenna case can be easily analyzed in a similar way.

\begin{figure}[htbp]
    \centering
    \includegraphics[width=0.68\linewidth]{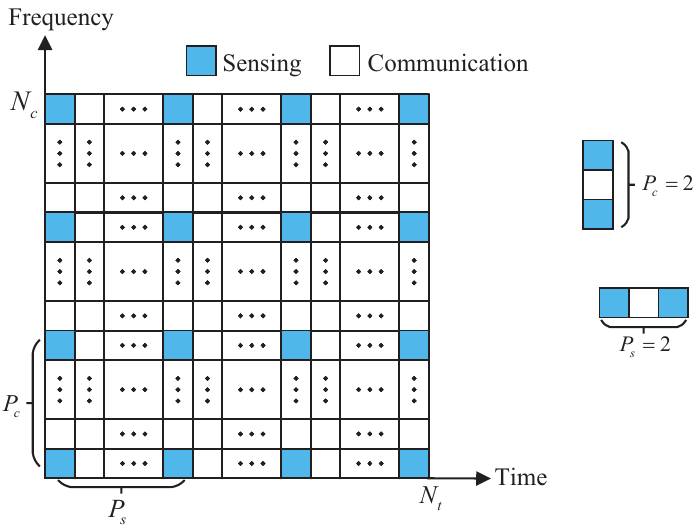}
    \caption{The distribution of RS on 2D time-frequency resources grid.}
    \label{fig2}
\end{figure}

In order to meet the various requirements of communication and sensing in ISAC systems, we design the waveform strategies by reconstructing the RS signal structure based on the OFDM frame. As shown in Fig. \ref{fig2}, the RS in a coherent processing interval (CPI) is uniformly inserted into a two-dimension (2D) time-frequency domain in a rectangular discrete manner, which composes of $N_{ t}$ OFDM symbols and $N_{ c}$ subcarriers. Here, $M$ OFDM symbols and $N$ subcarriers are for sensing with $1 < M \le N_{ t}$ and $1 < N \le N_{ c}$, and the rest for communications. For simplicity, each unit is denoted as a resource element (RE): each blue RE  $s_p(n,m)$ represents the modulated symbol for sensing over the $n$-th subcarry and the $m$-th OFDM symbol with $n \in \mathcal{N} = {\left \{ 0, 1, \cdots, N-1 \right \} }$ and $m \in \mathcal{M} = {\left \{ 0, 1, \cdots, M-1 \right \} }$; each blank RE $s_d(k,w)$ represents the modulated transmitted random data for communications over the $k$-th subcarry and the $w$-th OFDM symbol for communications, with $k \in \mathcal{K} =  {\left \{ 0, 1, \cdots, N_{ c} -N-1 \right \} }$ and $w \in \mathcal{W} =  {\left \{ 0, 1, \cdots, N_{ t} -M-1 \right \} }$. The intervals of blue REs in the frequency and time domains are denoted as $P_{ c} $ and  $P_{ s} $, respectively, with $P_c=\left \lceil \frac{N_c}{N}  \right \rceil  $ and $P_s=\left \lceil \frac{N_t}{M}  \right \rceil$ , and $\left \lceil x \right \rceil $ is a function of rounding a given number $x$ up to its nearest integer. Based on the above setup, we aim to achieve performance analysis between the sensing and communications by designing the flexible RS, with respect to parameters $P_{ c} $ and $P_{ s} $ in this paper. 

By the inverse discrete Fourier transform (IDFT), the modulated symbols in a CPI are transformed to the time domain \cite{liyanaarachchi_optimized_2021}. In the $t$-th slot, $t \in \mathcal{T} = {\left \{ 0, 1, \cdots, T-1 \right \} }$, the transmit signal at the BS is
\begin{align}
\label{2.1} 
x(t) = & \sum_{k \in \mathcal{K}}^{}\sum_{w \in \mathcal{W}}^{}s_d(k,w)e^{j2\pi k\Delta ft} rect(\frac{t-wT_{s}}{T_{s}})\\ \nonumber 
&+\sum_{n \in \mathcal{N} }^{}\sum_{m \in \mathcal{M}}^{}s_p(n,m)e^{j2\pi nP_{c}\Delta ft}rect(\frac{t-mP_{s}T_{s}}{T_{s}}), 
\end{align}
where the first term is for communication, and the second one is for sensing; $\Delta f$ is the subcarrier spacing; $T_{s}$ is the duration of each symbol; and $rect(\cdot )$ is the rectangle function.

\section{ISAC Performance Analysis} 
In this section, the sensing and communication performance are analyzed in our considered ISAC system, respectively.

\begin{figure*}[!b]
\hrulefill
\begin{equation}
\label{3.25}
\text{CRB}_{R}(P_{c} ,P_{s}) = \frac{I_{v,v}}{\alpha}= \frac{3 c^2(2N_{{t}}-P_{{s}}) P_{{c}}^3 P_{{s}}}{(A\pi )^2 \Delta f^2 \gamma (N_{{c}}-P_{{c}})N_{{c}}N_{{t}} \left ((N_{{t}}+P_{{s}})(N_{{c}}+P_{{c}})+6(N_{{t}}N_{{c}}-P_{{c}} P_{{s}}) \right ) }  , \tag{13}
\end{equation}

\begin{equation}
\label{3.26}
\text{CRB}_{v}(P_{c} ,P_{s})  = \frac{I_{R,R}}{\alpha}= \frac{3 c^2(2N_{{c}}-P_{{c}}) P_{{s}}^3 P_{{c}}}{(A\pi )^2 (T_{{s}} f_{{c}})^2 \gamma (N_{{t}}-P_{{s}})N_{{c}}N_{{t}} \left ((N_{{t}}+P_{{s}})(N_{{c}}+P_{{c}})+6(N_{{t}}N_{{c}}-P_{{c}} P_{{s}}) \right ) }  . \tag{14}
\end{equation}
\end{figure*}

\subsection{Sensing Performance Analysis}
In the $t$-th time slot, the BS sends signal $x(t)$ given in (\ref{2.1}) to the receiver. The receiver distinguishes the RS and the communication data according to their respective locations within the 2D time-frequency resources grid. The RS is extracted from the received signal for sensing according to its known location, which corresponds to the second term in (\ref{2.1}). Thus, the received signal at the BS is expressed as \cite{keskin_mimo-ofdm_2021}
\begin{align}\nonumber
	y_{_{s}}(t) = &A\sum_{n \in \mathcal{N} }^{}\sum_{m \in \mathcal{M}}^{}s_p(n,m)e^{-j2\pi nP_{{c}}\Delta ft} \\  \label{3.1.2} 
	&  e^{j2\pi mP_{{s}}T_{{s}}f_{{D}}}rect(\frac{t-mP_{{s}}T_{{s}}-\tau}{T_{{s}}})  +n_s(t)  \\ \nonumber
	= & A\sum_{n \in \mathcal{N} }^{}\sum_{m \in \mathcal{M}}^{}s_p(n,m)e^{-j2\pi nP_{{c}}\Delta ft} \\   \label{3.1.3} 
	&  e^{j2\pi mP_{{s}}T_{{s}}\frac{2\nu f_{c}} {c}}rect(\frac{t-mP_{{s}}T_{{s}}-\frac{2R} {c}}{T_{{s}}})+n_s(t),  
\end{align} 
where $A$ is the channel gain, $n_s(t)$ is the circularly symmetric complex Gaussian (CSCG) noise with zero mean and variance $\sigma_s^{2}$, $f_{D}=\frac{2\nu f_{c}} {c}$ is the Doppler frequency shift caused by the relative motion of the user to the BS, and $\tau=\frac{2R} {c}$ is the transmission delay. Here, (\ref{3.1.3}) is obtained by substituting $f_{D}=\frac{2\nu f_{c}} {c}$ and $\tau=\frac{2R} {c}$ into (\ref{3.1.2}), with $c$ being the velocity of light, $f_{c}$ being the central carrier frequency, $R$ and $v$ being the distance and velocity of the user in relation to the BS, respectively.

Based on the received signal $y_{_{s}}(t)$ in (\ref{3.1.3}), the BS aims to estimate distance and velocity of the target for sensing. Let $\boldsymbol{\theta} =[R,v]^T$ denote the estimation parameter vector. By discrete Fourier transform (DFT) and serial-to-parallel operations on $y_{_{s}}(t)$ , the modulated version of $y_{_{s}}(t)$ over the $n$-th subcarry and the $m$-th OFDM symbol, denoted as $y_s(n,m)$, is expressed as \cite{keskin_mimo-ofdm_2021}
	\begin{align} \label{3.17}  \nonumber
		y_s(n,m)=&As_{p}(n,m)e^{-j2\pi nP_{{c}}\Delta f \frac{2R} {c}}e^{j2\pi mP_{{s}}T_{{s}}\frac{2\nu f_{c}} {c}} \\
		&+n_s({n,m}),
	\end{align}
where noise $n_s({n,m})$ is converted from the $n_s(t)$ by DFT, and $rect(\cdot )$ in (\ref{3.1.3}) guaranteeing the orthogonality of the subcarriers in the time domain is removed due to it takes the value 1 for $t \in [0,1]$.

From (\ref{3.17}), $s_{p}(n,m)$ as a priori fixed symbol is known for sensing, and thus it can be removed from the received signal $y_s(n,m)$. Then, the removed version of $y_s(n,m)$, denoted as $z_{n,m}$, is given as
\begin{align}  \label{3.18}
z_{n,m}&=Ae^{-j2\pi nP_{{c}}\Delta f \frac{2R} {c}}e^{j2\pi mP_{{s}}T_{{s}}\frac{2\nu f_{c}} {c}}+\omega_{n,m}   \\ \label{3.188}
 &=s_{n,m} +\omega_{n,m},
\end{align}
where $s_{n,m}=Ae^{-j2\pi nP_{{c}}\Delta f \frac{2R} {c}}e^{j2\pi mP_{{s}}T_{{s}}\frac{2\nu f_{c}} {c}}$ and $\omega_{n,m}=\frac{\mathit{Var } (n_s({n,m}))}{ \left | s_{p}(n,m) \right |^2 }=\frac{\sigma_s^2}{{Q_{n,m}^2}}$, with $Q_{n,m}= \left | s_{p}(n,m) \right |  $ being the amplitude of the transmitted symbol and $\sigma_s^2$ being the variance of $n_s({n,m})$.

Based on (\ref{3.18}) and (\ref{3.188}), we aim to adopt the estimation $\boldsymbol{\theta}=[R,v]^T$ to derive the CRB matrix for sensing, which is defined as the inverse of Fisher information matrix (FIM). Hence, the FIM $\boldsymbol{\theta}$ is given as \cite{kay_fundamentals_1993} 
	\begin{align} 	\label{3.22}
		\boldsymbol{I_\theta} =\begin{bmatrix}I_{R,R}&I_{R,v}\\\\I_{v,R}&I_{v,v}\end{bmatrix} ,
	\end{align}
where each element in (\ref{3.22}) is
	\begin{align}
		\boldsymbol{I_\theta}(i,j) &=-E[\frac{\partial^2\ln p(z_{n,m};\boldsymbol{\theta})}{\partial{\theta _i}\partial{\theta _j}}]  \\ 	\label{3.23}
		&=\frac{Q_{n,m}^2}{2\sigma_s^2}\sum_{n=0}^{N-1}\sum_{m=0}^{M-1}\frac{\partial s_{n,m}}{\partial{\theta _i}}\frac{\partial s_{n,m}^\ast}{\partial{\theta _j}},
	\end{align}
with $i \in \left \{ {1, 2} \right \} $ and $j \in \left \{ {1, 2} \right \} $; $p(z_{n,m};\boldsymbol{\theta})$ being the conditional probability density function, i.e., 
\begin{align}
	p(z_{n,m};\boldsymbol{\theta}) & =\prod_{n=0}^{N-1}\prod_{m=0}^{M-1}\frac1{\sqrt{2\pi\frac{\sigma_s^2}{{Q^2_{n,m}}}}}e^{-\frac1{2\frac{\sigma_s^2}{{Q_{n,m}^2}}} q_{n,m}^2} \\ \label{3.20}
	& =\frac1{(2\pi\frac{\sigma_s^2}{{Q^2_{n,m}}})^{\frac{NM}2}}e^{-\frac{Q_{n,m}^2}{2\sigma_s^2}\sum_{n=0}^{N-1}\sum_{m=0}^{M-1}q_{n,m}^2},
\end{align}
and $q_{n,m} = \left | z_{n,m}-s_{n,m} \right |$. 

Then, the CRB matrix for estimation parameter $\boldsymbol{\theta} $ is given as 
\begin{align} \label{3.24}
\boldsymbol{I_\theta} ^{-1}= \frac{1}{\alpha} \begin{bmatrix}I_{v,v}&-I_{R,v}\\\\-I_{v,R}&I_{R,R}\end{bmatrix} ,
\end{align}
where $\alpha = I_{v,v}I_{R,R}-I_{v,R}I_{R,v}$.

\begin{Proposition}
\label{P2}
To simplify computation complexity, we express  $P_c=\left \lceil \frac{N_c}{N}  \right \rceil  $ and $P_s=\left \lceil \frac{N_t}{M}  \right \rceil$ as $P_c=\frac{N_c}{N}$ and $P_s=\frac{N_t}{M}$, respectively. Then, by calculating the values of $I_{v,v}$ and $I_{R,R}$ in matrix $\boldsymbol{I_\theta} ^{-1}$ in (\ref{3.24}), the closed-form expression of CRB for distance and velocity estimations, denoted as $\text{CRB}_{R}$ and $\text{CRB}_{v}$, are given in (\ref{3.25}) and (\ref{3.26}), respectively, where $\gamma = \frac{Q_{n,m}^2}{\sigma_s^2}$.
\end{Proposition}

From (\ref{3.25}) and (\ref{3.26}), it is observed that both the $\text{CRB}_{R}(P_{c} ,P_{s})$ and $\text{CRB}_{v}(P_{c} ,P_{s})$ are related to the intervals $P_{{c} }$ and $P_{{s}}$ of the RS in frequency and time domains. Intuitively, modeling the weighted sum of $\text{CRB}_{R}(P_{c} ,P_{s})$  and $\text{CRB}_{v}(P_{c} ,P_{s})$ in (\ref{3.25})-(\ref{3.26}) provides a comprehensive representation of the system's sensing performance \cite{jiang_rethinking_2022}, which is regarded as the performance metrics of both distance and velocity in our considered ISAC system, i.e.,
\setcounter{equation}{14}
\begin{align} \label{3.27}   \nonumber
\text{CRB}_{s}(P_{c} ,P_{s})=&\eta \text{CRB}_{R}(P_{{c} },P_{{s}}) \\
&+(1-\eta )\text{CRB}_{v}(P_{{c} },P_{{s}}), 
\end{align}
where $\eta\in(0,1)$ is the weight factor to balance the importance of distance and velocity estimations under different scenarios.  

\subsection{Communication Performance Analysis} 
The RS in (\ref{2.1}) is also utilized for channel estimation to enhance the communication performance. Specifically, the BS sends signal $x(t)$ given in (\ref{2.1}) to the user, and then the received signal is converted from the time domain to the frequency domain after undergoing DFT and serial-to-parallel operations at the user. Here, the transmitted modulation symbols $s_d(k,w)$ and $s_p(n,m)$ for both communication and sensing are uniformly expressed as $s(r,u)$, which is obtained by applying the DFT to $x(t)$ in time domain. Therefore, the received modulation symbol at the user is given as
	\begin{align}  \label{3.1}
		y_c(r,u) = h(r,u)*s(r,u)  +n(r,u),
	\end{align}
where $r=k+n$ and $u=w+m$ with $k$, $n$, $w$, and $m$ given in (\ref{2.1}); $h(r,u)$ is the channel coefficient from BS to the user over the $r$-th subcarry and the $u$-th OFDM symbol, with $r \in \mathcal{R} = {\left \{ 0, 1, \cdots, N_c-1 \right \} }$, $u \in \mathcal{U} = {\left \{ 0, 1, \cdots, N_t-1 \right \} }$; and $n(r,u)$ is the CSCG noise with zero mean and variance $\sigma ^2$, respectively. Considering a dual-selective channel and the channel is time invariant with an OFDM symbol, the $h(r,u)$ is expressed as \cite{choi_optimum_2005}  
	\begin{align} \label{3.2}
		h(r,u)=\sum_{l=0}^{L-1}h_l(uT_{s})e^{-j2\pi r \Delta f \tau_l},
	\end{align} 
where $L$ is the number of channel paths, and $h_l$ and $\tau_l$ denote the attenuation factor and the delay of the $l$-th path, respectively. Here, the paths are statistically independent and have normalized average gain, i.e., $\sum_{l=0}^{L-1}\sigma_l^2=1$, with $\sigma_l^2$ being the average gain of the $l$-th path.

From the RS part in (\ref{3.1}), the user performs channel estimation using the least squares (LS) method. Then, the interpolation is used to obtain the channel estimation for the remaining data locations in the 2D time-frequency resource. Oversampling the LS estimation by $P_{s}$ and $P_{c}$ factor in the time and frequency domains, respectively, $\mathbf{H}\in \mathbb{C} ^{N_{ c}\times N_{ t}}$ is perfectly estimated by ideal interpolator $\mathbf{W}_{p}$, i.e., 
	\begin{align} \label{3.3}
		\mathbf{H}=\mathbf{H}_{s} \mathbf{W}_{p},
	\end{align}
where $\mathbf{H}$ is the channel matrix with each element being $h(r,u)$ given in (\ref{3.2}); $\mathbf{H}_{ {s}}$ represents the channel matrix of RS and is equal to $\mathbf{H}$ for the ideal interpolator; $\mathbf{W}_{p} \in \mathbb{C} ^{N_{ c}\times N_{ t}}$ is the ideal 2D brick-wall-type noncausal filter without noise, with each element $w_p(nP_{c},mP_{s})$ being
	\begin{align} \label{3.4}
		w_p(nP_{c},mP_{s})=\frac{\text{sin}(\frac{\pi n}{P_{c}})\text{sin}(\frac{\pi m}{P_{s}})}{(\frac{\pi n}{P_{c}})(\frac{\pi m}{P_{s}})} ,nP_{c} \in \mathcal{R},mP_{s} \in \mathcal{U}.
	\end{align}
	
As the real LS estimation contains noisy samples of the channel, the channel estimation $\mathbf{\hat{H}}$ under nonideal interpolator $\mathbf{W}$ is 
	\begin{align} 	\label{3.5}
		\mathbf{\hat{H}}  =\mathbf{H}'_{ {LS}} \mathbf{W},
	\end{align}
where $\mathbf{H}'_{ {LS}}$ is the real LS estimation of $\mathbf{H}$, i.e., 
	\begin{align} \label{3.6}
			\mathbf{H}'_{ {LS}}=\mathbf{H}_{s}+\mathbf{N}_{s},
	\end{align}
with $\mathbf{N}_{s}$ being the CSCG noise with zero mean and variance $\mathbf{N} _s\sim \mathcal{CN} (0,\sigma_N^{2}\mathbf{I} )$, with $\mathbf{I} \in \mathbb{C} ^{N_{ c}\times N_{ t}}$ being the identity matrix. By substituting (\ref{3.6}) into (\ref{3.5}), the channel estimation $\mathbf{\hat{H}}$ is rewritten as 
\begin{align} \label{3.7}
	\mathbf{\hat{H} } &=  (\mathbf{H}_{s}+\mathbf{N}_{s}) \mathbf{W} \\ \label{3.76}
	&= (\mathbf{H}_{s}+\mathbf{N}_{s}) \mathbf{W}+\mathbf{H}-\mathbf{H}_{s} \mathbf{W}_{p} \\  \label{3.77}
	&= \mathbf{H}+\mathbf{H}_{s} (\mathbf{W}-\mathbf{W}_{p})+\mathbf{N}_{s}\mathbf{W}, 
\end{align}
where (\ref{3.76}) is obtained by (\ref{3.3}) and (\ref{3.7}); (\ref{3.77}) is obtained by rearranging terms in (\ref{3.76}). Then, the estimation error between $\mathbf{\hat{H}}$ and $\mathbf{H}$ is expressed as 
\begin{align} \label{3.8}
	\mathbf{e = \hat{H} - H}=\mathbf{H}_{s} \mathbf{W}_e + \mathbf{N}_{s}\mathbf{W} ,
\end{align}
where $\mathbf{W}_e=\mathbf{W}-\mathbf{W}_{p}$; $\mathbf{N}_{s}$ is a sampled version of $\mathbf{N}$ with density $D = (P_{s}P_{c})^{-1}$; $\mathbf{N}$ is the noise matrix with each element being $n(r,u)$ given in (\ref{3.1}), and we have $\sigma_N^{2}=\sigma^{2}D$.

After 2D DFT and discrete Parseval's theorem for the estimation error $\mathbf{e}$ in (\ref{3.8}), the MSE of channel estimation for each element of matrix $\mathbf{e}$ is written as\cite{ribeiro_impact_2007} 
\begin{align} \label{3.9}  \nonumber
\sigma_{e}^{2} =& \frac1{4\pi^{2}}\int_{-\pi}^{\pi}\int_{-\pi}^{\pi}\left|H_{s}(w_{n},w_{m})\right|^{2}{\left|W_{e}(w_{n},w_{m})\right|}^{2}dw_{n}dw_{m} \\
&+\frac{\sigma^{2} D}{4\pi^{2}}\int_{-\pi}^{\pi}\int_{-\pi}^{\pi}\left|W(w_{n},w_{m})\right|^{2}dw_{n}dw_{m},
\end{align}
where $\left|H_{s}(w_{n},w_{m})\right|^{2}$ is a sampled version of power spectral density $S_{H}(w_{n},w_{m})$ of the original channel, i.e., 
\begin{align}  \label{3.10}
 \left|H_{s}(w_{n},w_{m})\right|^{2} = \frac{1}{(P_{ s}P_{ c})^2} S_{H}(w_{n},w_{m}),
\end{align}
and the error filter $W_{e}(w_{n},w_{m})$ of 2D FFT in $\left \{ \left | w_n \right | \le \pi / P_{ c}, \left | w_m \right | \le \pi / P_{ s} \right \} $ is written as 
\begin{align} \label{3.11}
W_{e}(w_{n},w_{m}) = W(w_{n},w_{m}) -  P_{ s} P_{ c},
\end{align}
with linear interpolator $W(w_{n},w_{m})$ being represented as
\begin{align} \label{3.12}
W(w_n,w_m)=\frac{1}{P_{ s}P_{ c}}\left[\frac{\sin\left(\frac{P_{ c}w_n}{2}\right)}{\sin\left(\frac{w_n}{2}\right)}\frac{\sin\left(\frac{P_{ s}w_m}{2}\right)}{\sin\left(\frac{w_m}{2}\right)}\right]^2.
\end{align}

To obtain the explicit representation of ${\sigma}_{e}^{2}$ with respect to $P_{ s}$ and $P_{ c}$, (\ref{3.9}) is approximated by the second-order Taylor series of $W_{e}(w_{n},w_{m}) $ in (\ref{3.12}), i.e., 
\begin{align} \label{3.13}
{\sigma}_{e}^{2}  \approx {\sigma}_{e}'^{2} = \frac{2P_{ s}^2P_{ c}^2\widehat{w}_{n} ^2\widehat{w}_{m} ^2+P_{ c}^4\widehat{w}_{n} ^4+P_{ s}^4\widehat{w}_{m} ^4}{144} + \frac{P_{ s}P_{ c} \sigma^{2}}{4\pi^{2}} ,
\end{align}
where $\widehat{w}_{n} ^z=\frac{1}{2\pi } \int_{-\pi }^{\pi } w_{n} ^z S_{H}(w_{n})dw_{n}$ and $\widehat{w}_{m} ^z=\frac{1}{2\pi } \int_{-\pi }^{\pi } w_{m} ^z S_{H}(w_{m})dw_{m}$ are the $z$-th order moment of the Doppler spectrum and power delay profile, respectively.

From (\ref{3.13}), it is observed that the channel estimation error decreases with $P_{ s}$ and $P_{ c}$ decreasing, which may result in better communication performance. On the other hand, the communication performance is also affected by the number of REs for communications. In a CPI, The communication rate in our considered ISAC system is represented as \cite{ohno_capacity_2004} 
\begin{align} \label{3.14}
C(P_{c} ,P_{s})={\Delta f}\sum_{k \in \mathcal{K}}^{}\sum_{w \in \mathcal{W}}^{}{\log} \left(1+\frac{\sigma_l^2 \sigma_{\widehat{H} }^2}{\sigma_l^2{\sigma}_e^2+\sigma^2}\right),
\end{align}
where $\sigma_{\widehat{H} }^2$ is the variance of the estimator in frequency domain.
\begin{Proposition}
\label{P1}
By substituting (\ref{3.13}) into (\ref{3.14}), the communication rate is rewritten as
\begin{align} \label{3.15} 
C'(P_{c} ,&P_{s}) =  {\Delta f}(N_{ c}N_{ t}- \frac{N_{ c}N_{ t}}{P_{ c}P_{ s}})    \\   \nonumber
& {\log} \left(1+\frac{\sigma_{\widehat{H} }^2}{\frac{1}{144}(P_{ s}^2\widehat{w}_{m} ^2+P_{ c}^2\widehat{w}_{n} ^2)^2+ (\frac{P_{ c}P_{ s} }{4\pi^{2}} +1 )\frac{\sigma^2}{\sigma_l^2} }\right).     
\end{align}
\end{Proposition}  

\begin{Remark}
From (\ref{3.15}), it is observed that the communication rate $C'(P_{c} ,P_{s})$ decreases with the intervals $P_{ s}$ and $P_{ c}$ decreasing. Oppositely, the channel estimation ${\sigma}_{e}^{2}$ becomes more precise. Therefore, it is necessary to investigate the effects of the intervals $P_{ s}$ and $P_{ c}$ on the communication performance. 
\end{Remark}

\section{ISAC Problem and Formulation}
Our goal is to minimize the weighted sum CRB of distance and velocity estimations, i.e., $\text{CRB}_s(P_{c} ,P_{s})$ given in (\ref{3.27}), by designing the intervals $P_{c}$ and $P_{s}$, subject to communication rate constraint. Then, the corresponding optimization problem is formulated as
\begin{align} 
\label{4.1} 	\min_{\{ P_{c} ,P_{s}\}} ~~~~&~\text{CRB}_s(P_{c} ,P_{s})\\
	\label{4.2}\textrm{s.t}.~~~~~~&~C(P_{c} ,P_{s}) \ge  C_{min},\\ \label{4.33}
	~~~&~P_{{c} }\in \left \{1, 2,\cdots , N_{{c}}  -1 \right \}  , \\ 
	\label{4.3}~~~&~P_{{s} }\in \left \{1, 2,\cdots ,  N_{{t}}-1  \right \},  
\end{align} 
where (\ref{4.2}) is the communication rate constraint, with $C_{min}$ representing the minimum communication requirement; (\ref{4.3}) and (\ref{4.33}) are the constraints for intervals $P_{c}$ and $P_{s}$, respectively.

\begin{Remark}
It is observed that problem (\ref{4.1})-(\ref{4.3}) is a mixed-integer problem due to the discreted values $P_{c}$ and $P_{s}$, which is intractable in general. The intuitive method is to exhaustive search the optimal values for $P_{c}$ and $P_{s}$. However, as the numbers of OFDM symbols and subcarriers $N_{ c}$ and $N_{ t}$ increase, the computational complexity grows exponentially.
\end{Remark}

To address the above issue, both the values $P_{ s}$ and $P_{ c}$ are regard as continuous variables, and the original problem (\ref{4.1})-(\ref{4.3}) is transformed into
\begin{align}
\label{4.11} 	\min_{\{ P_{c} ,P_{s}\}} ~~~~&~\text{CRB}_s(P_{c} ,P_{s})\\  \label{4.22}   
\textrm{s.t}.~~~~~~&~~  C_{min} - C(P_{c} ,P_{s}) \le 0 ,\\  
\label{4.333}~~~~&~~  1-P_{{c} }  \le  0,1- P_{{s} }\le  0 , \\
\label{4.44}~~~~&~~P_{{c} }- N_{{c}}  +1 \leq 0, P_{{s} }- N_{{t}}+1 \leq 0,
\end{align}  
where (\ref{4.22}) is obtained by rearranging terms in (\ref{4.2}); (\ref{4.333}) and (\ref{4.44}) are the continuous versions obtained by  (\ref{4.33}) and (\ref{4.3}), respectively. 

It is easy to observe that problem (\ref{4.11})-(\ref{4.44}) is non-convex, which is difficult to be solved. Therefore, an approximation method is proposed to solve this problem in this paper. First, within a CPI, there must be $N_{{c}}\gg P_{{c}}$ and $N_{{t}}\gg P_{{s}}$. Then, the weighted sum CRB of distance and velocity estimations in (\ref{4.11}) is approximated as

	\begin{align} \label{4.6} \nonumber
		\text{CRB}_s (P_{c} ,P_{s})\approx  & \eta \frac{6 c^2 P_{{c}}^3 P_{{s}}}{7(A\pi )^2 \Delta f^2 \gamma N_{{c}}^3N_{{t}} } + \\
		& (1-\eta ) \frac{6 c^2 P_{{s}}^3 P_{{c}}}{7(A\pi )^2 (T_{{s}} f_{{c}})^2 \gamma N_{{t}}^3 N_{{c}} } .
	\end{align}   
It is obvious that (\ref{4.6}) is convex.
	
Employing $C'(P_{c} ,P_{s})$ in (\ref{3.15}) instead of $C(P_{c} ,P_{s})$ in (\ref{4.22}) facilitates the establishment of a straightforward relationship between the communication rate and intervals. Thus, we reformulate the constraint $C_{min} - C'(P_{c} ,P_{s}) \le 0$ into a standard form, 
\begin{align} \label{4.4}
	& {\frac{C_{min}{P_{ c}P_{ s}}}{ {\Delta f}N_{ c}N_{ t}({P_{ c}P_{ s}}- 1)}}-  \\ \nonumber
	&{\log} \left(1+\frac{\sigma_{\widehat{H} }^2}{\frac{1}{144}(P_{ s}^2\widehat{w}_{m} ^2+P_{ c}^2\widehat{w}_{n} ^2)^2+ (\frac{P_{ c}P_{ s} }{4\pi^{2}} +1 )\frac{\sigma^2}{\sigma_l^2} }\right) \le 0.
\end{align}
From (\ref{4.4}), it is observed that the first term is a convex function with respect to $P_{c}$ and $P_{s}$, and the second term is a concave function with respect to $P_{c}$ and $P_{s}$ according to the composite rule. The subtraction of a convex function and a concave function is a convex function, and thus (\ref{4.4}) is convex.

Based on the above analysis, the original problem (\ref{4.1})-(\ref{4.3}) is transformed into
\begin{align}
\label{4.7} 	\min_{\{ P_{c} ,P_{s}\}} ~~~~&~\text{CRB}_s(P_{c} ,P_{s})\\   
\label{4.8} \textrm{s.t}.~~~~~~&~~\text{(\ref{4.333}),(\ref{4.44}),(\ref{4.4})} .
\end{align}  
It is easy to observe that the objective function and constraints in problem (\ref{4.7})-(\ref{4.8}) are convex. Hence, it can be efficiently solved by some optimization tools, e.g., CVX. 

\section{Simulations and Numerical Results}   
In this section, simulation results are provided to validate our analysis. The carrier frequency is set as $f_{c}=$28GHz, and the frequency spacing of subcarriers is set as $\Delta f = $120KHz. The total duration of OFDM symbal is set as $T_s=\text{8.92}\mu $s, and the bandwidth is set as 100MHz. In a CPI, the number of subcarriers and OFDM symbols are set as $N_{{c}}=$792 and $N_{{t}}=$448, respectively, and $P_{c}$ and $P_{s}$ are set in the range of $[2,15]$. 

\begin{figure}[htbp]
	\centering
	\includegraphics[width=0.86\linewidth]{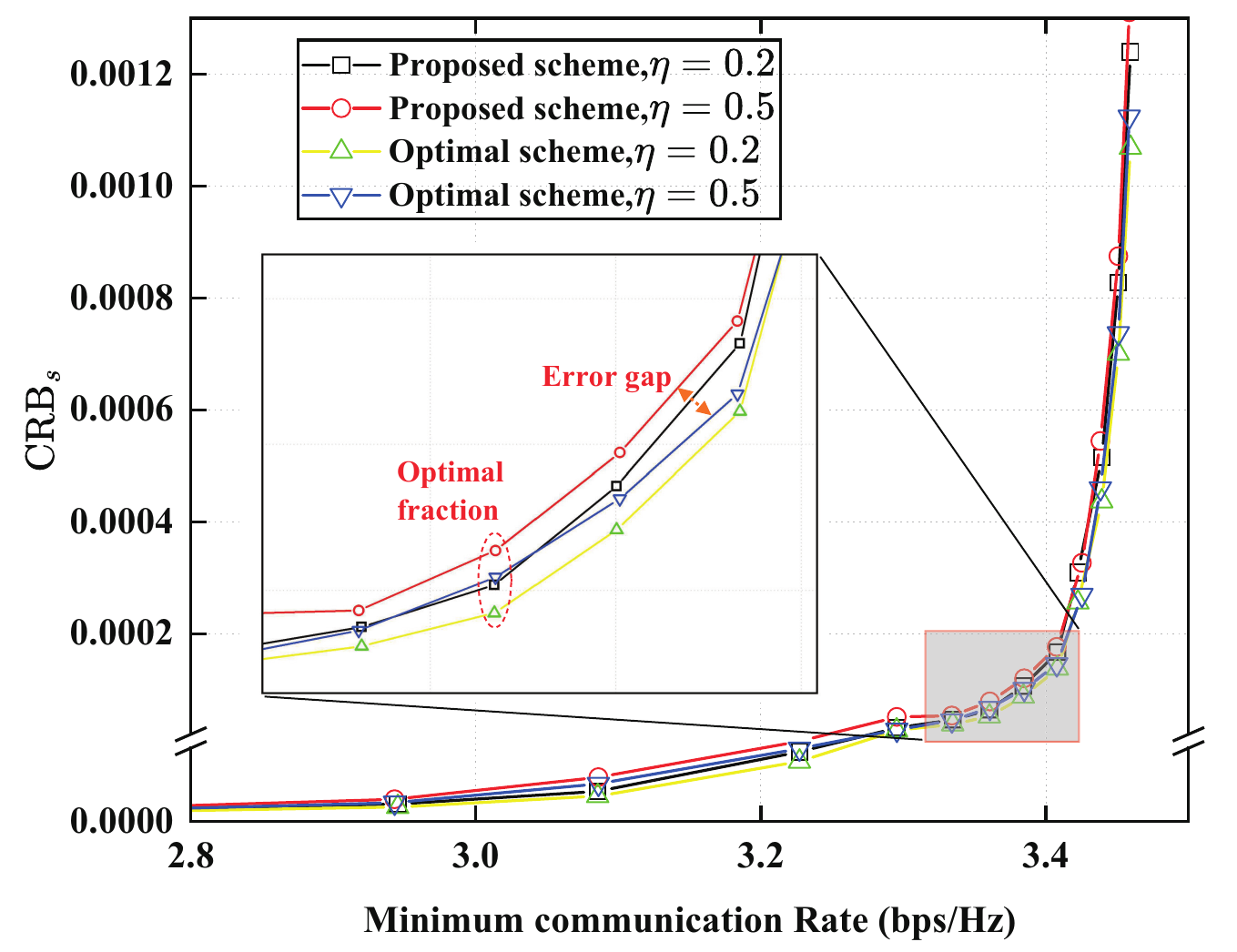}
	\caption{Trade-off curve between $\text{CRB}_s$ and minimum communication rate with different schemes and $\eta$.}
	\label{fig3}
\end{figure}

\begin{figure}[htbp]
	\centering
	\includegraphics[width=0.86\linewidth]{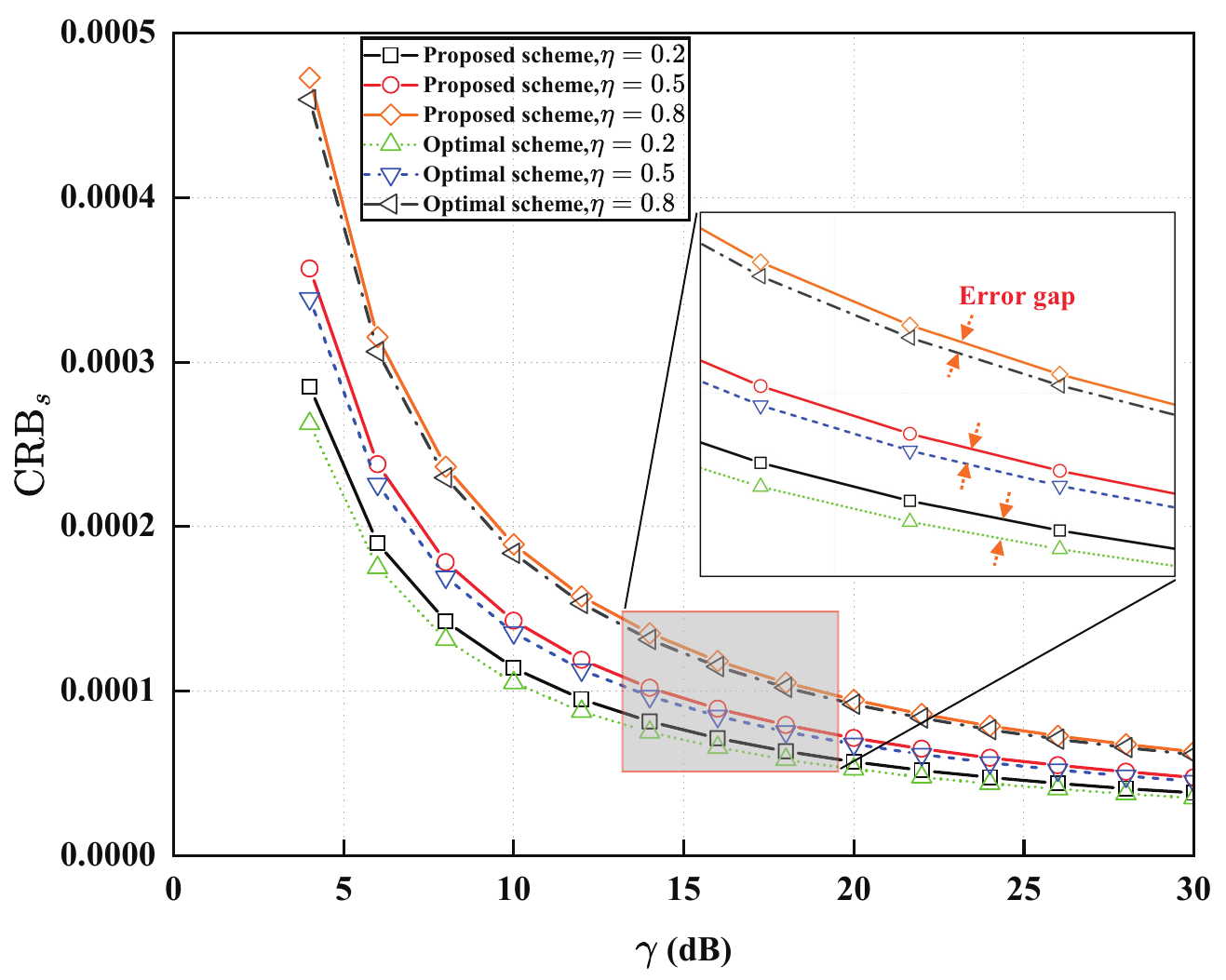}
	\caption{The impact of the $\gamma$ on the $\text{CRB}_s$ with different schemes and $\eta$.}
	\label{fig4}
\end{figure}

Fig. \ref{fig3} shows the trade-off curve between $\text{CRB}_s$ and minimum communication rate of our proposed scheme and the exhaustive search scheme under the different $\eta$. It is observed that the sensing performance significantly decreases as the minimum communication rate increases. This is due to the fact that more number of REs are used for communications. We can also see the optimal fraction for the performance of communication and sensing. When $\eta= 0.2$, the optimal fraction takes $P_{c}=6$ and $P_{s}=4$. When $\eta= 0.5$, the optimal fraction takes $P_{c}=4$ and $P_{s}=6$. This is because $\text{CRB}_{R}$ is more related to interval $P_{c}$ in the frequency domain and the $\text{CRB}_{v}$ is more related to interval $P_{s}$ in the time domain. Moreover, it shows the error gap between the proposed sub-optimal and the optimal exhaustive search scheme and the error gap does not affect the optimal fraction. Then, Fig. \ref{fig4} shows the impact of $\gamma$ on the $\text{CRB}_s$, with different $\eta$ under the proposed scheme and exhaustive search method, with $P_{c}$ and $P_{s}$ of the optimal fraction. It is observed that there is error gap between the proposed scheme and exhaustive search method due to the relaxation, and this gap further decreases with large weight factor.

\section{Conclusion}     
This paper considered a waveform strategy by designing the RS in OFDM-based ISAC system. The CRB for distance and velocity estimations by RS for sensing were derived in a closed-form and the communication rate was also obtained under the MSE of channel estimation. Then, a weighted sum CRB minimization problem was formulated under the communication rate and RS intervals constraints. A relaxation and approximation method was proposed to solve this problem. Finally, numerical results verified that our analysis, and showed slight gap of our proposed scheme compared to the exhaustive search method.

\bibliography{ref}

\begin{thebibliography}{10}
\providecommand{\url}[1]{#1}
\csname url@samestyle\endcsname
\providecommand{\newblock}{\relax}
\providecommand{\bibinfo}[2]{#2}
\providecommand{\BIBentrySTDinterwordspacing}{\spaceskip=0pt\relax}
\providecommand{\BIBentryALTinterwordstretchfactor}{4}
\providecommand{\BIBentryALTinterwordspacing}{\spaceskip=\fontdimen2\font plus
\BIBentryALTinterwordstretchfactor\fontdimen3\font minus
  \fontdimen4\font\relax}
\providecommand{\BIBforeignlanguage}[2]{{%
\expandafter\ifx\csname l@#1\endcsname\relax
\typeout{** WARNING: IEEEtran.bst: No hyphenation pattern has been}%
\typeout{** loaded for the language `#1'. Using the pattern for}%
\typeout{** the default language instead.}%
\else
\language=\csname l@#1\endcsname
\fi
#2}}
\providecommand{\BIBdecl}{\relax}
\BIBdecl

\bibitem{wang_integrated_2022}
J.~Wang, N.~Varshney, C.~Gentile, S.~Blandino, J.~Chuang, and N.~Golmie,
  ``\BIBforeignlanguage{en}{Integrated {Sensing} and {Communication}:
  {Enabling} {Techniques}, {Applications}, {Tools} and {Data} {Sets},
  {Standardization}, and {Future} {Directions}},''
  \emph{\BIBforeignlanguage{en}{IEEE Internet Things J.}}, vol.~9, no.~23, pp.
  23\,416--23\,440, Dec. 2022.

\bibitem{wang_receiver_2024}
\BIBentryALTinterwordspacing
D.~Wang, Y.~Tian, C.~Huang, H.~Chen, X.~Xu, and P.~Zhang,
  ``\BIBforeignlanguage{en}{Receiver {Selection} and {Transmit} {Beamforming}
  for {Multi}-static {Integrated} {Sensing} and {Communications}},'' Jul. 2024.
  [Online]. Available: \url{http://arxiv.org/abs/2407.05873.}
\BIBentrySTDinterwordspacing

\bibitem{tian_performance_2024}
Y.~Tian, D.~Wang, C.~Huang, and W.~Zhang, ``\BIBforeignlanguage{en}{Performance
  {Trade}-off of {Integrated} {Sensing} and {Communications} for {Multi}-{User}
  {Backscatter} {Systems}},'' \emph{\BIBforeignlanguage{en}{IEEE Trans.
  Wireless Commun.}}, Sep. 2024, Early access.

\bibitem{hua_mimo_2024}
H.~Hua, T.~X. Han, and J.~Xu, ``\BIBforeignlanguage{en}{{MIMO} {Integrated}
  {Sensing} and {Communication}: {CRB}-{Rate} {Tradeoff}},''
  \emph{\BIBforeignlanguage{en}{IEEE Trans. Wireless Commun.}}, vol.~23, no.~4,
  pp. 2839--2854, Apr. 2024.

\bibitem{liyanaarachchi_optimized_2021}
S.~D. Liyanaarachchi, T.~Riihonen, C.~B. Barneto, and M.~Valkama,
  ``\BIBforeignlanguage{en}{Optimized {Waveforms} for {5G}–{6G}
  {Communication} {With} {Sensing}: {Theory}, {Simulations} and
  {Experiments}},'' \emph{\BIBforeignlanguage{en}{IEEE Trans. Wireless
  Commun.}}, vol.~20, no.~12, pp. 8301--8315, Dec. 2021.

\bibitem{ma_frac_2021}
D.~Ma, N.~Shlezinger, T.~Huang, Y.~Liu, and Y.~C. Eldar,
  ``\BIBforeignlanguage{en}{{FRaC}: {FMCW}-{Based} {Joint}
  {Radar}-{Communications} {System} {Via} {Index} {Modulation}},''
  \emph{\BIBforeignlanguage{en}{IEEE J. Sel. Top. Signal Process.}}, vol.~15,
  no.~6, pp. 1348--1364, Nov. 2021.

\bibitem{liu_integrated_2022}
F.~Liu, Y.~Cui, C.~Masouros, J.~Xu, T.~X. Han, Y.~C. Eldar, and S.~Buzzi,
  ``\BIBforeignlanguage{en}{Integrated {Sensing} and {Communications}: {Toward}
  {Dual}-{Functional} {Wireless} {Networks} for {6G} and {Beyond}},''
  \emph{\BIBforeignlanguage{en}{IEEE J. Sel. Areas Commun.}}, vol.~40, no.~6,
  pp. 1728--1767, Jun. 2022.

\bibitem{wei_integrated_2023}
Z.~Wei, H.~Qu, Y.~Wang, X.~Yuan, H.~Wu, Y.~Du, K.~Han, N.~Zhang, and Z.~Feng,
  ``\BIBforeignlanguage{en}{Integrated {Sensing} and {Communication} {Signals}
  {Toward} {5G}-{A} and {6G}: {A} {Survey}},''
  \emph{\BIBforeignlanguage{en}{IEEE Internet Things J.}}, vol.~10, no.~13, pp.
  11\,068--11\,092, Jul. 2023.

\bibitem{chen_joint_2021}
L.~Chen, F.~Liu, W.~Wang, and C.~Masouros, ``\BIBforeignlanguage{en}{Joint
  {Radar}-{Communication} {Transmission}: {A} {Generalized} {Pareto}
  {Optimization} {Framework}},'' \emph{\BIBforeignlanguage{en}{IEEE Trans.
  Signal Process.}}, vol.~69, no.~5, pp. 2752--2765, May 2021.

\bibitem{li_frame_2024}
Y.~Li, F.~Liu, Z.~Du, W.~Yuan, Q.~Shi, and C.~Masouros,
  ``\BIBforeignlanguage{en}{Frame {Structure} and {Protocol} {Design} for
  {Sensing}-{Assisted} {NR}-{V2X} {Communications}},''
  \emph{\BIBforeignlanguage{en}{IEEE. Trans. Mob. Comput.}}, Apr. 2024, Early
  access.

\bibitem{hsu_analysis_2022}
H.-W. Hsu, M.-C. Lee, M.-X. Gu, Y.-C. Lin, and T.-S. Lee,
  ``\BIBforeignlanguage{en}{Analysis and {Design} for {Pilot} {Power}
  {Allocation} and {Placement} in {OFDM} {Based} {Integrated} {Radar} and
  {Communication} in {Automobile} {Systems}},''
  \emph{\BIBforeignlanguage{en}{IEEE Trans. Veh. Technol.}}, vol.~71, no.~2,
  pp. 1519--1535, Feb. 2022.

\bibitem{ji_networking_2023}
K.~Ji, Q.~Zhang, Z.~Wei, Z.~Feng, and P.~Zhang,
  ``\BIBforeignlanguage{en}{Networking {Based} {ISAC} {Hardware} {Testbed} and
  {Performance} {Evaluation}},'' \emph{\BIBforeignlanguage{en}{IEEE Commun.
  Mag.}}, vol.~61, no.~5, pp. 76--82, May 2023.

\bibitem{cui_integrated_2022}
Y.~Cui, X.~Jing, and J.~Mu, ``\BIBforeignlanguage{en}{Integrated {Sensing} and
  {Communications} {Via} {5G} {NR} {Waveform}: {Performance} {Analysis}},'' in
  \emph{\BIBforeignlanguage{en}{Proc. IEEE Int. Conf. Acoust. Speech Signal
  Process.(ICASSP)}}, Singapore, Singapore, May 2022, pp. 8747--8751.

\bibitem{wei_5g_2023}
Z.~Wei, Y.~Wang, L.~Ma, S.~Yang, Z.~Feng, C.~Pan, Q.~Zhang, Y.~Wang, H.~Wu, and
  P.~Zhang, ``\BIBforeignlanguage{en}{{5G} {PRS}-{Based} {Sensing}: {A}
  {Sensing} {Reference} {Signal} {Approach} for {Joint} {Sensing} and
  {Communication} {System}},'' \emph{\BIBforeignlanguage{en}{IEEE Trans. Veh.
  Technol.}}, vol.~72, no.~3, pp. 3250--3263, Mar. 2023.

\bibitem{keskin_mimo-ofdm_2021}
M.~F. Keskin, H.~Wymeersch, and V.~Koivunen,
  ``\BIBforeignlanguage{en}{{MIMO}-{OFDM} {Joint} {Radar}-{Communications}:
  {Is} {ICI} {Friend} or {Foe}?}'' \emph{\BIBforeignlanguage{en}{IEEE J. Sel.
  Top. Signal Process.}}, vol.~15, no.~6, pp. 1393--1408, Nov. 2021.

\bibitem{kay_fundamentals_1993}
S.~M. Kay, \emph{\BIBforeignlanguage{en}{Fundamentals of statistical signal
  processing: estimation theory}}.\hskip 1em plus 0.5em minus 0.4em\relax USA:
  Prentice-Hall, Inc., Feb. 1993.

\bibitem{jiang_rethinking_2022}
J.~Jiang, M.~Xu, Z.~Zhao, K.~Han, Y.~Li, Y.~Du, and Z.~Wang,
  ``\BIBforeignlanguage{en}{{Rethinking} the {Performance} of {ISAC} {System}:
  {From} {Efficiency} and {Utility} {Perspectives}},'' in
  \emph{\BIBforeignlanguage{en}{Proc. ACM MobiCom Workshop Integr. Sens.
  Commun. Syst., Part MobiCom.(ISACom)}}, New York, USA, Oct. 2022, pp. 19--24.

\bibitem{choi_optimum_2005}
J.-W. Choi and Y.-H. Lee, ``\BIBforeignlanguage{en}{Optimum {Pilot} {Pattern}
  for {Channel} {Estimation} in {OFDM} {Systems}},''
  \emph{\BIBforeignlanguage{en}{IEEE Trans. Wireless Commun.}}, vol.~4, no.~5,
  pp. 2083--2088, Sep. 2005.

\bibitem{ribeiro_impact_2007}
C.~Ribeiro and A.~Gameiro, ``\BIBforeignlanguage{en}{On the {Impact} of the
  {Pilot} {Density} in the {Channel} {Estimation} of {MC}-{CDMA} {Systems}},''
  in \emph{\BIBforeignlanguage{en}{Proc. IEEE Int. Symp. Wirel. Commun.
  Syst.(ISWCS)}}, Trondheim, Norway, Oct. 2007, pp. 777--781.

\bibitem{ohno_capacity_2004}
S.~Ohno and G.~Giannakis, ``\BIBforeignlanguage{en}{Capacity {Maximizing}
  {MMSE}-{Optimal} {Pilots} for {Wireless} {OFDM} over {Frequency-}{Selective}
  {Block} {Rayleigh}-fading {Channels}},'' \emph{\BIBforeignlanguage{en}{IEEE
  Trans. Inf. Theory}}, vol.~50, no.~9, pp. 2138--2145, Sep. 2004.

\end{thebibliography}

\end{document}